# An Exploration of the Impact of Mapping Style and Device Roadmap on Simulated ReRAM Architectures for Neuromorphic Computing

by

Enrico F. Persico

University of Illinois at Urbana-Champaign

July 2023

## Abstract

This paper investigates the relationship between mapping style and device roadmap in Resistive Random Access Memory (ReRAM) architectures for neuromorphic computing. The study leverages simulations using DNN+NeuroSim to evaluate the impact of different parameters on chip performance, including latency, energy consumption, and overall system efficiency. The results demonstrate that novel mapping techniques and a high-performance (HP) device roadmap are optimal if energy and speed considerations are weighted equally. This is because as the study demonstrates, HP devices provide a latency cut that outsizes the energy cost. Additionally, adopting novel mapping in the device cuts latency by nearly 30% while being slightly more energy efficient. The findings highlight the importance of considering mapping style and device roadmap in optimizing ReRAM architectures for neuromorphic computing, which may contribute to advancing the practical implementation of ReRAM in computational systems.

* I would like to thank Dr. Xuanyao Fong for his guidance and mentorship throughout this research process and term. I would also like to thank Yunuo Cen for the technical help he so generously provided throughout said term.

# Table of Contents

# List of Tables and Figures

# 1. Introduction

Emerging non-volatile memories (eNVMs) are revolutionizing the field of memory research, offering immense potential for transformative advances in computing. These novel memory technologies combine the low latency of RAM with the permanent storage capabilities of SSD, presenting exciting prospects for reducing power consumption in large-scale computational systems. Among the various eNVM technologies, Resistive Random Access Memory (ReRAM) has garnered significant attention due to its unique characteristics and promising applications in neuromorphic computing.

ReRAM is based on memristors, a breakthrough innovation introduced by a team at HP Labs in 2008. Memristors store resistance values on a dielectric, typically utilizing a p-doped film of titanium dioxide adjacent to a non-doped film. Applying a charge to the combined slab causes oxygen ions to migrate to the doped film, facilitating increased electron flow and resulting in reduced resistance. These memristor cells are arranged in a crossbar structure, enabling access through standard RAM addressing (word line/bit line) mechanisms (Hayes 2018).

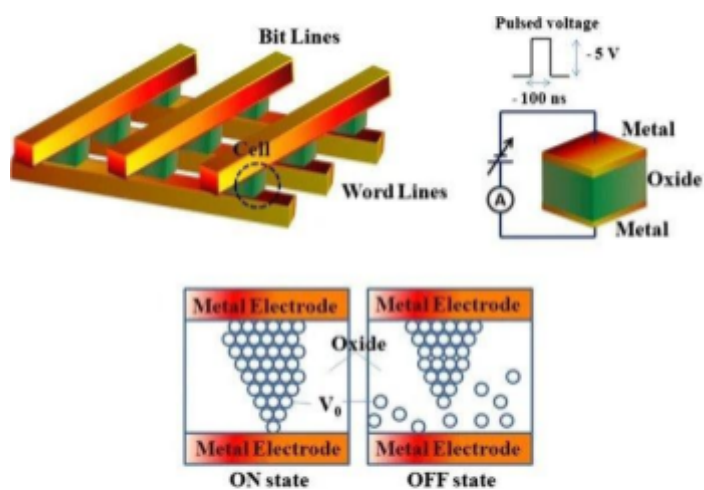

Figure 1 - A visual depiction of the fundamental structures of ReRAM: crossbar array (top left), simplified circuit diagram (top right), and chemical diagram (bottom). Source: Meena 2014.

While ReRAM holds tremendous promise, its implementation faces challenges related to heat susceptibility. The extreme temperatures common in modern data centers, which could greatly benefit from ReRAM technology, pose significant issues. High temperatures induce resistance drift and elevated current leakage in ReRAM cells, compromising data integrity and overall reliability.

Nevertheless, ReRAM offers substantial potential in diverse applications, particularly in the realm of neuromorphic computing. In this context, ReRAM and other eNVMs offer exceptional speed and power efficiency, which can greatly benefit emerging technologies like Internet of Things (IoT) devices, autonomous vehicles, and robotic devices. Exploring the optimal configuration of ReRAM memory systems is crucial to harnessing their full potential, making it essential to investigate the impact of design changes on simulated performance.

Therefore, this study focuses on investigating the relationship between mapping style and device roadmap and their influence on chip performance in ReRAM architectures. By leveraging simulations conducted using DNN+NeuroSim, we aim to gain insights into the potential of ReRAM in neuromorphic computing applications. This research explores various parameters, such as mapping style, device roadmap, bit size, operation mode, pipeline configuration, buffer type, buffer size, and chip activation, to understand their impact on latency, energy consumption, and overall performance. The findings from this study will contribute to optimizing chip design and advancing the practical implementation of ReRAM in various computational systems.

## 2. Simulated Architectures And Their Properties

This study leverages simulations conducted using DNN+NeuroSim, which is a sophisticated simulation tool employed in this study to evaluate the performance of eNVM-based hardware architectures. It offers a comprehensive framework for training and analyzing simulated hardware architectures using an image recognition dataset. By utilizing DNN+NeuroSim, we can gain valuable insights into the potential of ReRAM in image recognition and related applications.

The simulations conducted with DNN+NeuroSim involve training a simulated eNVM-based hardware architecture on the image recognition dataset CIFAR-10. This process includes mapping the neural network computations onto the hardware resources, configuring the device roadmap, and specifying other parameters such as bit size, operation mode, pipeline configuration, buffer type, buffer size, and chip activation (Peng, Dec. 2020). The simulated hardware architecture emulates the behavior of a real-world eNVM-based system.

During the simulations, the performance of the hardware architecture is monitored and recorded. This includes measuring latency, energy consumption, and overall chip performance. The recorded data provides insights into how the selected parameters, such as mapping style and device roadmap, affect the performance of the ReRAM architecture in image recognition tasks. By analyzing the results, we can understand the advantages and limitations of ReRAM in comparison to other memory technologies for these specific applications.

Additionally, the simulations with DNN+NeuroSim enable evaluation of the impact of different design choices on the performance of ReRAM-based hardware architectures. It allows for exploration of trade-offs between latency, energy consumption, and overall system efficiency. By leveraging the capabilities of DNN+NeuroSim, researchers can optimize the configuration and design of ReRAM architectures, ultimately advancing their practical implementation in various computational systems.

The customization of the simulation is mainly done in the "Param.cpp" file. The following list demonstrates the functionality of each of the main parameters and what impact this may have on the architecture:

> **Mapping (Conventional/Novel):**
>
> Mapping refers to the way neural network computations are assigned and distributed across the hardware resources. In DNN+NeuroSim, mapping can be either conventional or novel. Conventional mapping refers to the traditional approach of mapping neural network kernels to the crossbar by mapping each kernel to a column in the crossbar structure. The novel mapping technique instead maps kernels into submatrices and allocates them in different areas in the memory. This results in higher memory usage (see Table 1 for details), which improves efficiency in NeuroSim. That's because NeuroSim only uses the simulated memory for one task, meaning that higher memory usage is unilaterally good within the context of the simulation. However, it should be noted that in other applications in which the memory needs to be used for different tasks, lower memory utilization from one task can be beneficial.

Table I Memory Utilization

| Network | Conventional Mapping | Novel Mapping |
|---|---|---|
| VGG-8 (CIFAR-10) | 91.45% | 95.23% |
| AlexNet | 98% | 97% |
| VGG-16 | 98.79% | 99.24% |
| ResNet-34 | 85.88% | 90.13% |

Source: Peng, Mar. 2020

**Device Roadmap (HP v. LSTP):**

Device roadmap refers to the specific technology or design approach used for the memory devices in the hardware architecture. In DNN+NeuroSim, two commonly considered device roadmaps are HP (High Performance) and LSTP (Low Standby Power). These roadmaps represent different tradeoffs between performance and power consumption, and choosing one over the other can impact the overall performance and energy efficiency of the neural network system.

**Bit Size:**

Bit size refers to the number of bits used to represent the values in the memory cells of the hardware architecture. In DNN+NeuroSim, the bit size parameter determines the precision and dynamic range of the data stored and processed in the memory cells. It is a critical factor that impacts the accuracy and efficiency of the neural network model, as well as the signal error on a hardware level.

# 3. Results

This study focuses on investigating the relationship between mapping style and device roadmap and their impact on chip performance. The results were obtained through the utilization of DNN+NeuroSim, simulated on the Feynman computing cluster at the Computational Nanoelectronics & Nanodevices Lab at the National University of Singapore. To view the raw summary data obtained from the simulations, you can see it on my GitHub (URL: https://github.com/enricopersico/neurosim-results).

Latency and energy consumption simulations were conducted using different device roadmaps and mapping styles, as depicted in Figures 2 and 3. The following key findings were observed:

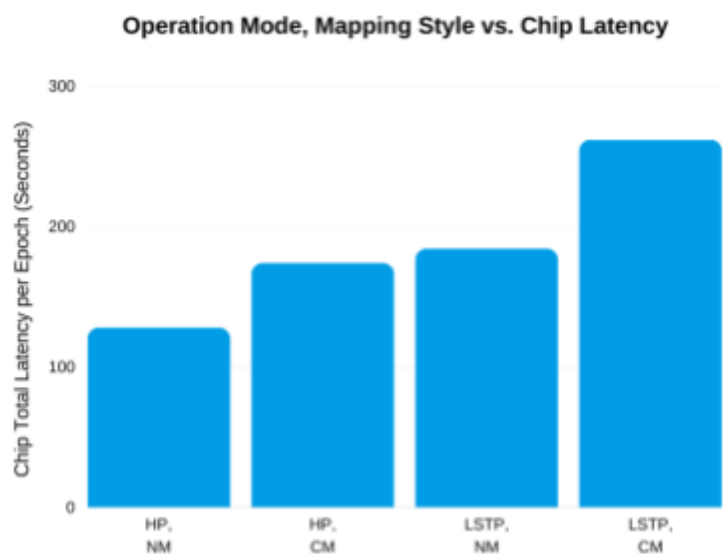

Figure 2.

**The Impact of Mapping Style and Device Roadmap on Energy Consumption:**

Switching the device roadmap from Low Standby Power (LSTP) to High Performance (HP) resulted in an average decrease in latency of approximately 32%. This indicates that the adoption of an HP device roadmap significantly improves chip performance in terms of latency.

In addition, switching from Conventional Mapping (CM) to Novel Mapping (NM) led to an average reduction in latency of ~28%. This finding demonstrates that the utilization of novel mapping techniques can effectively reduce latency and enhance chip performance.

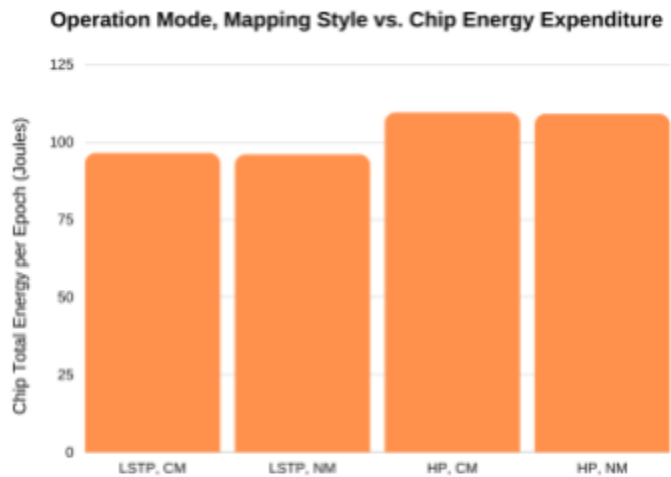

Figure 3.

**The Impact of Mapping Style and Device Roadmap on Energy Consumption:**

While novel mapping exhibited slightly higher energy efficiency, the overall effect of mapping style on energy expenditure was found to be negligible. The difference in energy consumption between CM and NM was only around 0.4%. Thus, the choice of mapping style appears to have a minimal impact on energy efficiency.

In contrast to mapping style, changing the device roadmap from HP to LSTP showed a notable impact on energy efficiency. The average decrease in energy consumption observed when switching from HP to LSTP was approximately 12%, from around 110 joules total to around 95. This indicates that adopting an LSTP device roadmap can significantly improve energy efficiency in chip performance.

Determining the worthiness of a tradeoff can be a challenging task. To aid in this decision-making process, the following formula has been proposed:

$$Score = \frac{1}{1000} \cdot Latency \cdot Energy$$

This formula assigns equal importance to both latency and energy, implying that neither factor holds more significance than the other. The following table illustrates how the model grades the combinations of mapping style and device roadmap in question:

Table II Efficiency Scoring

| Combination | Total Latency per Epoch (Seconds) | Total Energy per Epoch (Joules) | Score |
| --- | --- | --- | --- |
| LSTP, CM | 261.773 | 96.343 | 25.22 |
| HP, CM | 173.896 | 109.404 | 19.02 |
| LSTP, NM | 184.272 | 95.899 | 17.67 |
| HP, NM | 127.955 | 108.964 | 13.94 |

Interestingly, the model indicates that altering the mapping would yield greater benefits compared to modifying the device roadmap. Furthermore, despite the compromise in energy efficiency, the formula strongly suggests that adopting the HP roadmap is undoubtedly worthwhile.

These findings highlight the importance of considering both mapping style and device roadmap in optimizing chip performance. The adoption of novel mapping techniques and an HP device roadmap can lead to substantial reductions in latency, while the choice of device roadmap also significantly affects energy efficiency. Future research should further explore the implications of different combinations of mapping style and device roadmap to maximize chip performance and energy efficiency.

# 4. Discussion

While the findings are intriguing and offer valuable insights, it is important to acknowledge the limitations and potential factors that may impact the accuracy and generalizability of the results. The following list outlines some of these limitations:

**Relationship Measurement Parameters:**

One factor to consider is that the relationships examined in this study were measured using a default set of parameters. It is worth noting that altering these parameters might not fundamentally change the observed trends, but it could significantly impact the magnitude of the changes. For instance, employing the pipeline process in the simulated system could potentially alter the extent to which latency is affected by a differing mapping style.

**Assumption of an Ideal Two-Terminal Selector Device:**

In the simulated eNVM crossbar array, this study assumed the presence of an ideal two-terminal selector device. While this assumption allows for exploration of certain aspects of the system's behavior, it may not fully represent the real-world conditions or constraints of the technology being investigated. The use of an idealized device could potentially affect the accuracy and applicability of the results obtained.

**Optimistic Results for Higher Bit-Count Systems:**

Throughout the study, a constant bit count of 4 was maintained across all experiments. However, it is essential to highlight a limitation associated with the simulation, which yielded optimistic results for higher bit-count systems. This optimism in the simulation's outcomes can be attributed to factors such as simplified modeling assumptions and inadequate consideration of potential sources of error, including resistance drift and the increased impact of electrical noise as the bit count rises. Consequently, it is crucial to exercise caution when extrapolating these findings to real-world systems characterized by higher bit counts.

Despite the aforementioned limitations, the results obtained from this study offer valuable insights that contribute to the existing knowledge in the field. The identified limitations present opportunities for future research to address and mitigate these shortcomings, allowing for a more comprehensive understanding of the system's behavior. Future studies could explore the effects of varying measurement parameters on the observed relationships, providing deeper insights into the magnitude and dynamics of changes. Additionally, the creation of more realistic device models with the eNVM crossbar array can enhance the accuracy and applicability of the results, providing a better understanding of the technology's performance in practical scenarios.

## 5. Conclusions

The findings of this study reveal several significant trends regarding the impact of mapping style and device roadmap on latency and power consumption in ReRAM architectures. The results demonstrate that both novel mapping techniques and the adoption of a high-performance (HP) device roadmap can substantially reduce latency by nearly 30% each. Additionally, the study suggests that novel mapping exhibits superior performance, significantly reducing latency while marginally decreasing power consumption. Furthermore, the efficiency scoring model suggests that HP architectures are more efficient despite the ~12% energy cut offered.

This question highlights the need for further investigation and consideration in the design and optimization of ReRAM devices. Future research should aim to strike a balance between achieving low latency and minimizing power consumption to meet the diverse requirements of various applications. By exploring innovative approaches, such as combining novel mapping techniques with energy-efficient device roadmaps, memory designers can strive to achieve both improved performance and reduced energy consumption in ReRAM architectures.